# Prediction of a Two-dimensional Phosphorus Nitride Monolayer

Hang Xiao[1], Feng Hao[1], Xiangbiao Liao[1], Xiaoyang Shi[1] Yayun Zhang[1,3] and Xi Chen[1,2]

[1] *Columbia Nanomechanics Research Center, Department of Earth and Environmental Engineering, Columbia University, New York, NY 10027, USA*

[2] *SV Laboratory, School of Aerospace, Xi'an Jiaotong University, Xi'an 710049, China*

[3] *College of Power Engineering, Chongqing University, Chongqing 400030, China*

Today, 2D semiconductor materials have been extended into the nitrogen group: phosphorene, arsenene, antimonene and even nitrogene. Motivated by them, based upon first-principles density functional calculations, we propose a new two-dimensional phosphorus nitride (PN) structure that is stable well above the room temperature, due to its extremely high cohesive energy. Unlike phosphorene, PN structure is resistant to high temperature oxidation. The structure is predicted to be a semiconductor with a wide, indirect band gap of 2.64 eV. More interestingly, the phosphorus nitride monolayer experiences an indirect-to-direct band-gap transition at a relatively small tensile strain. Such dramatic transformation in the electronic structure combined with structural stability and oxidation resistance at high temperature could pave the way for exciting innovations in high-speed ultrathin transistors, power electronic modules, ultra-high efficiency LEDs and semiconductor lasers.

which enables its application in next generation of power electronics and clean energy technologies.

October 2004 marked the discovery of graphene [1], the first stable and truly 2D material. This epic discovery has opened up the possibility of isolating and studying the intriguing properties of a whole family of 2D materials including the 2D insulator boron nitride (BN) [2–4], 2D semiconductor molybdenum disulfide [5–7] and very recently, 2D phosphorus, i.e. phosphorene [8], which extend the 2D material family into the nitrogen group.

Unlike graphene, phosphorene has an inherent, direct band gap [9]. With its high carrier mobility up to ~ 1000 $cm^2 V^{-1} s^{-1}$ and an on/off ratio of ~ $10^4$ at ambient temperature [9], phosphorene is considered to be a very promising material for future applications in electronics and optoelectronics. However, the Achille's heel of phosphorene for practical applications is its instability in air [10–13]. Recently, using ab initio calculations, Danil W. Boukhvalov predicted that a small amount of nitrogen impurities are able to increase the chemical inertness of phosphorene without significant changes to its electronic structure [14]. If a slight doping of nitrogen in phosphorene could stabilize it, what if we dope it with nitrogen by 50%?

In this work, we propose a new binary counterpart of phosphorene consisting of equal number of nitrogen and phosphorene atoms in puckered configuration with space group Pmn21, as shown in Fig. 1(a). Our prediction of PN structure is obtained from first-principles density functional calculations with Perdew–Burke–Ernzerhof (PBE) [15] exchange-correlation functional. All our ab initio calculations were performed using the Cambridge series of total-energy package (CASTEP) [16,17]. A plane-wave cutoff energy of 520 eV is used, and a 9 ×6 ×1 Monkhorst-Pack [18] *k*-point mesh was used. The convergence test of cutoff energy and *k*-point mesh has been conducted. Because the band gaps are dramatically underestimated by the GGA level DFT [19,20], band structures of PN structure were calculated with HSE06 [21] hybrid functional, which has been demonstrated to be able to predict accurate band structures and density of states (DOS) [22]. All structure optimizations were conducted without imposing any symmetry constraints. The conjugate gradient method (CG) was used to optimize the atomic positions until the change in total energy was less than 5 ×$10^{-6}$ eV/atom, and the maximum displacement of atoms was less than 5 ×$10^{-5}$ Å.

The fully optimized the single layer PN structure is shown in Fig. 1(a). It is observed that the puckered structure is slightly deformed compared to the structure of phosphorene. Despite the structure similarity between the PN monolayer and phosphorene, the P-N bond in PN monolayer is much stronger than the P-P bond in phosphorene, since the cohesive energy of PN structure is 1.26 eV larger than that of phosphorene (Table 1). This indicates the stability of the PN structure. The unit cell (inset to Fig. 1(a)) consists of four atoms with lattice constants $a$ = 2.68 Å, $b$ = 4.17 Å. In the puckered structure, there are two types of P-N bond with bond lengths $d_1$ = 1.80 Å and $d_2$ = 1.71 Å and four types of bond angles: $\theta_1$ = 97.8 °, $\theta_2$ = 124.4 ° and $\theta_3 = \theta_4$ = 103.1 °(illustrated in Fig. 1(a)).

By conducting phonon dispersion relation calculation of the PN structure, we verify that all of phonon frequencies are real (Fig. 1(c)), indicating its structural stability.

To further test the stability of the structure even at high temperature with or without oxygen atmosphere, *ab initio* molecular dynamics (MD) simulations (shown in Fig. 2) at the PBE [15] /GTH-DZVP [23] level with NVT ensembles of the CP2K [24] code. The simulations were run for 10 ps at 1300 K, 1400K and at 1300 K in the oxygen atmosphere, respectively. For the simulation in the oxygen atmosphere, spin-polarized ab initio MD calculation is used to correctly model the triplet state of $O_2$. The stability of PN structure is maintained at 1300 K. However, at 1400 K, the crystalline structure dissociates into one dimensional N-P chains. These MD calculations indicates the stability of PN structure well above the room temperature. Intriguingly, the PN structure is even stable in the oxygen atmosphere at 1300 K for 10 ps, manifesting its oxidation resistance, which phosphorene lacks.

The band structures of unstrained and strained PN structure are shown in Fig. 3 and the Brillouin zone with the relevant high-symmetry k-points are depicted in Fig. 1(b). Calculations carried out by HSE06 show that the unstrained PN structure is a semiconductor with a wide indirect band gap of 2.64 eV. This is a well-sought characteristic, since all 2D semiconductors formed by first- and second-row elements reported thus far exhibit band gaps that are smaller than 2 eV. Intriguingly, the PN structure becomes a direct bandgap semiconductor with band gap of 2.68 eV, when 4% tensile strain in the zigzag direction is applied. We also computed the effective mass of the electron (shown in Fig. 3) for the strained PN structure at the Γ point along the Γ-X (armchair) and the Γ–Y (zigzag) directions. The quadratic band dispersion near the conduction band minimum is assumed when we estimate the electron effective mass. The effective electron masses were found to be $m_e^{\Gamma Y} = 0.31\ m_o$ and $m_e^{\Gamma X} = 1.05\ m_o$, where $m_o$ is the free-electron mass. The electron effective mass along the Γ –Y (zigzag) direction is much lighter than that along the Γ-X (armchair) direction, showing a strong anisotropic transport property. The carrier mobility is expected to high due to its small electron effective mass. High carrier mobility combined with wide direct band gap enable its future applications in high-speed ultrathin transistors, ultra-high efficiency light-emitting diodes and semiconductor lasers. Furthermore, the band gap of PN structure can be modulated by stacking or rolling up to form PN nanotubes, opening up even more possibilities for its potential application.

In conclusion, we have presented a prediction of a new two-dimensional phosphorus nitride material with distinguished structures and outstanding properties. Band structures calculated using HSE06 hybrid functional indicate that monolayered phosphorus nitride is indirect semiconductor with wide band gap of 2.64 eV. Intriguingly, monolayered phosphorus nitride exhibits an indirect-to-direct band-gap transition at a relatively small tensile strain. Such dramatic transformation in the electronic structure combined with structural stability and oxidation resistance at high temperature could pave the way for exciting innovations in high-speed ultrathin transistors, power electronic modules, ultra-high efficiency LEDs and semiconductor lasers.

## FIGURES AND TABLES

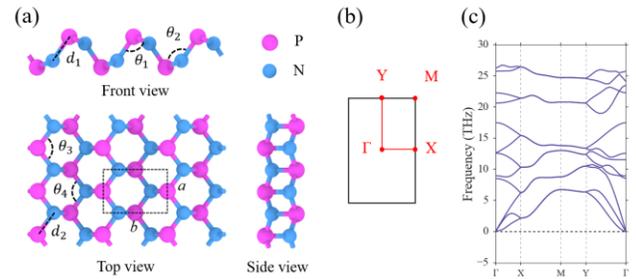

Fig. 1. (a) 2D crystalline structure (top view, front view and side view) of the nitrogen phosphide with lattice constants $a$ = 2.68 Å, $b$ = 4.17 Å, bond lengths $d_1$ = 1.80 Å and $d_2$ = 1.71 Å, bond angles $\theta_1$ = 97.8 °, $\theta_2$ = 124.4 °and $\theta_3 = \theta_4$ = 103.1 °. (b) The Brillouin zone, with the relevant high-symmetry k points indicated. (c) The phonon dispersion relation for the PN monolayer. Structural stability is indicated by the absence of negative frequencies.

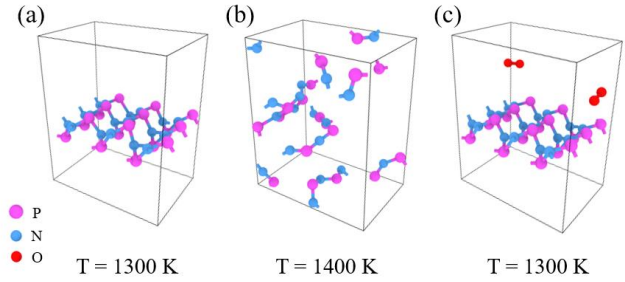

| (a) | (b) | (c) |

T = 1300 K    T = 1400 K    T = 1300 K

● P
● N
● O

Fig. 2. Snapshots of the atomic structure at (a) 1300 K, (b) 1400 K and at (c) 1300 K in oxygen atmosphere in the *Ab initio* MD simulation at time 10 ps. The stability of PN structure is maintained at 1300 K, while at 1400 K the crystalline structure dissociates into multiple N-P chains. Moreover, the PN structure is stable in oxygen atmosphere at 1300 K for 10 ps. These MD calculations indicates the stability of PN structure well above the room temperature.

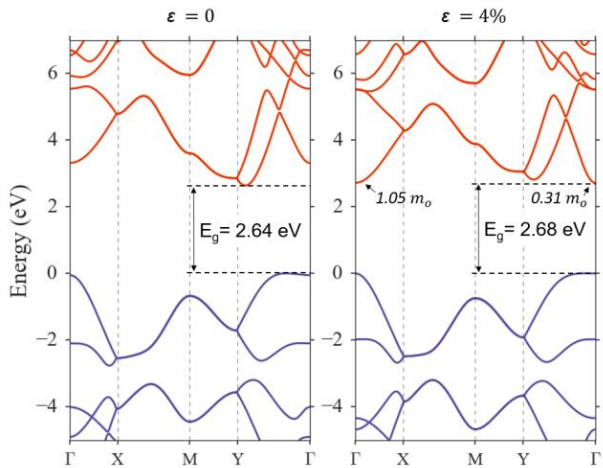

Fig. 3. Calculated band structure obtained with the HSE06 hybrid functional for the (a) strain-free PN unit cell (b) PN unit cell with tensile strain (applied in the zigzag direction), $\varepsilon = 4\%$. Strain-induced indirect to direct bandgap transition for PN structure is observed at $\varepsilon = 4\%$. The electron effective mass of for strained PN at the $\Gamma$ point along the $\Gamma$-X and the $\Gamma$-Y directions are indicated by black arrows.

| Structure | Cohesive energy (eV/ atom) |
|---|---|
| Phosphorene | 4.63 |
| PN monolayer | 5.89 |

Table. 1. Cohesive energies (at 0 K) using the PBE functional for two 2D structures, each consisting of 4 atoms unit cells. The cohesive energy for PN monolayer is calculated using the formula

$E_c = \frac{2E(N) + 2E(P) - E(PN\ unit\ cell)}{4}$.

## ACKNOWLEDGEMENTS

The study is supported by DARPA (W91CRB-11-C-0112), AFOSR (FA9550-12-1-0159) and the National Natural Science Foundation of China (11172231).